\documentclass[
 reprint,
 superscriptaddress,
 amsmath,amssymb,
 aps,
 prl,
 floatfix,
]{revtex4-2}

\usepackage[utf8]{inputenc}
\usepackage{graphicx}
\usepackage{dcolumn}
\usepackage{bm}
\usepackage{booktabs}

\begin{document}

\title{Confined Oxygen-Vacancy Migration Drives Ferroelectric Switching}

\author{Hyeonhu Bae}
\affiliation{Department of Condensed Matter Physics, Weizmann Institute of Science, Rehovot, 7610001, Israel}
\affiliation{Department of Physics, the Pennsylvania State University, University Park, PA, 16802, USA}
\author{Binghai Yan}
\email{binghai.yan@psu.edu}
\affiliation{Department of Condensed Matter Physics, Weizmann Institute of Science, Rehovot, 7610001, Israel}
\affiliation{Department of Physics, the Pennsylvania State University, University Park, PA, 16802, USA}
\date{\today}

\begin{abstract}
Conventional ferroelectricity arises from intrinsic lattice distortions, whereas oxygen vacancies are generally regarded as detrimental because their migration induces leakage currents and polarization degradation. However, the recently discovered ultrathin van der Waals ferroelectric Bi$_2$SeO$_5$ exhibits robust out-of-plane polarization switching despite its pristine crystal symmetry forbidding the corresponding displacive ferroelectric instability. Here we show that this apparent contradiction originates from confined oxygen-vacancy migration. We find that oxygen vacancies preferentially form within SeO$_3$ units and undergo reversible low-barrier rearrangements between nearly degenerate configurations. These localized vacancy dynamics generate a large switchable out-of-plane polarization, while long-range vacancy diffusion is suppressed by substantially higher migration barriers. At a representative vacancy concentration of 2.5\%, the resulting polarization reaches approximately $16~\mu\text{C}\,\text{cm}^{-2}$, consistent with experiment. Our results identify confined oxygen-vacancy migration as the microscopic origin of ferroelectric switching in Bi$_2$SeO$_5$ and establish defect-enabled ferroelectricity as a general mechanism for layered van der Waals oxides.
\end{abstract}

\maketitle

\begin{figure}
    \centering
    \includegraphics[width=0.85\columnwidth]{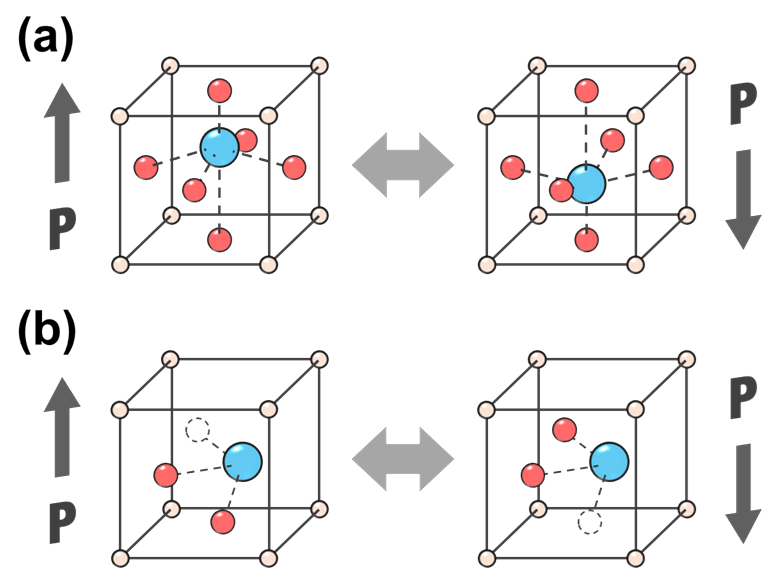}
    \caption{\textbf{Design principle for confined vacancy migration-driven ferroelectricity.}
    \textbf{a} Conventional ionic displacive switching in a perovskite oxide.
    \textbf{b} Vacancy-mediated switching, where a local oxygen vacancy rearrangement reverses the dipole while long-range diffusion is suppressed. The filled orange circle and open circle denote an oxygen ion and an oxygen vacancy, respectively.
    }
    \label{fig:schematic}
\end{figure}

Ferroelectric field-effect transistors (FeFETs) are promising for nonvolatile memory and low-power logic because reversible polarization in the gate dielectric can modulate channel charge without static power consumption~\cite{khan2020future}. Continued device scaling has therefore intensified interest in ultrathin ferroelectrics that remain switchable down to the nanometer and even unit-cell thickness limit~\cite{cheema2020enhanced,cheema2022emergent}. In the conventional picture, ferroelectricity arises from intrinsic lattice instabilities of the pristine crystal, with switching driven by reversible atomic displacements~\cite{dawber2005physics,xue2021emerging,taniguchi2013ferroelectricity}, as illustrated in Fig.~\ref{fig:schematic}a. Defects, especially oxygen vacancies (V$_\text{O}$), are usually viewed as detrimental in this framework because their migration causes leakage, imprint, internal bias fields, and fatigue, ultimately degrading device reliability~\cite{lee2023role,nukala2021reversible,rushchanskii2021ordering}.

This picture changes when vacancy motion is confined. If a vacancy can move only within a local structural unit, its motion can become reversible and polar-active rather than destructive (see Fig.~\ref{fig:schematic}b, for example). Layered van der Waals (vdW) oxides provide a natural platform for this behavior because anisotropic bonding and the vdW gap may suppress long-range vacancy migration while still allowing local rearrangements~\cite{wang2023towards,xue2021emerging,zhang2023single,li2020native}. A confined vacancy can then act as a switchable microscopic degree of freedom rather than a source of degradation.

\begin{figure*}[tb]
    \centering
    \includegraphics[width=0.99\textwidth]{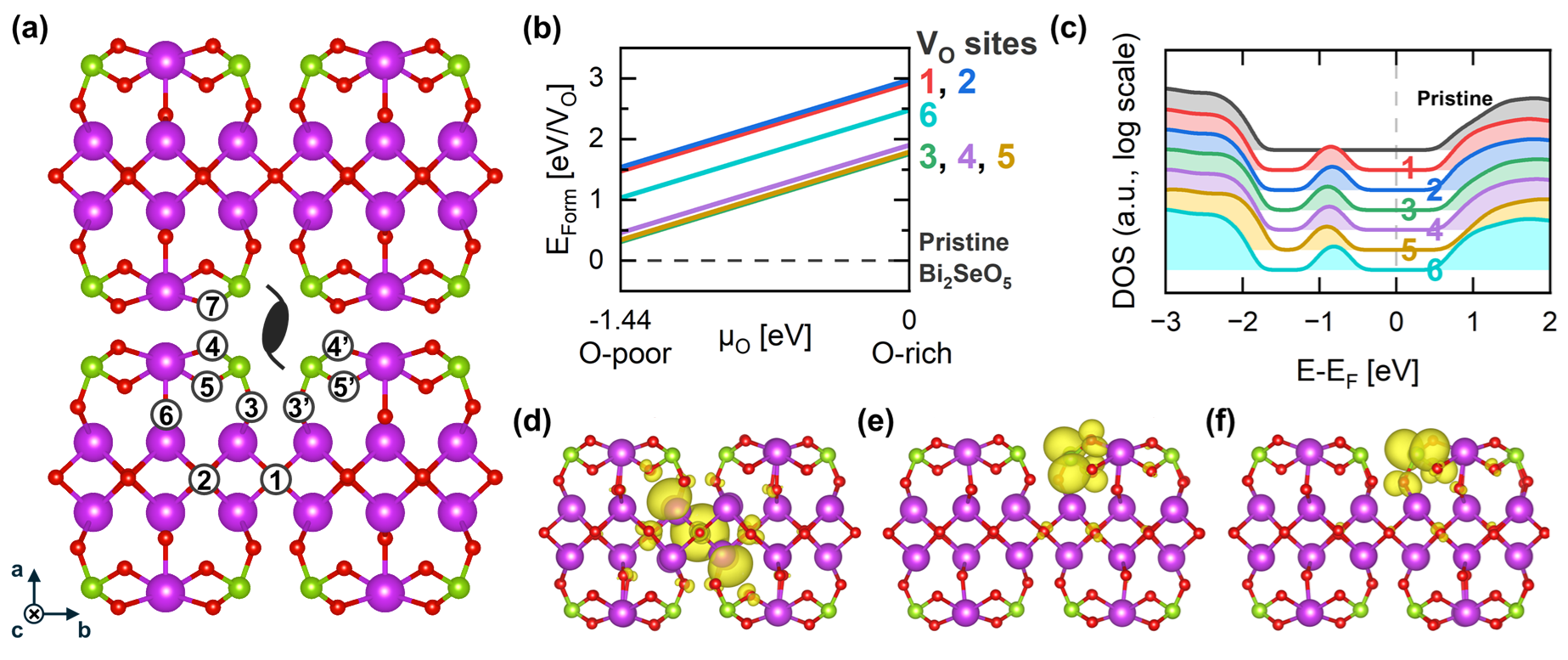}
    \caption{\textbf{Energetics of oxygen vacancy sites and localized defect states.}
    \textbf{a} Bi$_2$SeO$_5$ supercell with symmetry-inequivalent O sites 1--6, glide-related counterparts 3$'$--5$'$, and the nearest interlayer site 7. The twofold rotation axis that forbids $a$-axis polarization is shown at the center.
    \textbf{b} Oxygen vacancy (V$_\text{O}$) formation energies as a function of oxygen chemical potential, $E_\text{form}$($\mu_\text{O}$), showing that sites 3--5 are thermodynamically favored compared with other sites.
    \textbf{c} Density of states of vacancy-containing structures, showing an insulating in-gap defect level.
    \textbf{d-f} Charge densities of the in-gap state for vacancies at sites 1 (\textbf{d}), 3 (\textbf{e}), and 4 (\textbf{f}). The site 1 preserves the twofold symmetry, whereas sites 3 and 4 break it and localize the defect state on the neighboring Se-O network.
    }
    \label{fig:vacancysites}
\end{figure*}

Bi$_2$SeO$_5$ provides an especially compelling system in which to examine this possibility. It was recently discovered as a ferroelectric high-$\kappa$
dielectric oxide with out-of-plane switching down to the monolayer limit~\cite{wu2026wafer}. Yet this observation poses a microscopic puzzle. In pristine Bi$_2$SeO$_5$, the crystal symmetry allows the in-plane polarization but strictly forbids the existence of out-of-plane polarization (Fig.~\ref{fig:vacancysites}a). 
The microscopic origin of the out-of-plane ferroelectric switching cannot be explained by the intrinsic lattice symmetry, calling for an alternative mechanism other than the displacive switching. 

In this work, we propose that confined oxygen-vacancy migration leads to the ferroelectric properties of Bi$_2$SeO$_5$. Oxygen vacancies, which are charge neutral, undergo reversible low-barrier rearrangements inside the SeO$_3$ units in the crystal, generating a large switchable out-of-plane polarization. In contrast, long-range migration is suppressed by substantially higher barriers. At a representative 2.5\% vacancy concentration, the resulting polarization is estimated to be about $16~\mu\text{C}\,\text{cm}^{-2}$, comparable to the experimental report of $22~\mu\text{C}\,\text{cm}^{-2}$~\cite{wu2026wafer}. Our results therefore identify confined oxygen-vacancy migration as the microscopic origin of ferroelectric switching in Bi$_2$SeO$_5$ and suggest a new route to defect-driven ferroelectricity in layered vdW oxides.

\paragraph*{Thermodynamics of oxygen vacancies.}

Bi$_2$SeO$_5$ is a wide-gap insulator that crystallizes in the polar space group $Aem2$ (SG \#39)~\cite{rademacher2001crystal,dityatyev2004phase,li2020native,dong2024exploring}. The van der Waals layer corresponds to the $bc$-plane with $a$ being the out-of-plane direction, as shown in Fig.~\ref{fig:vacancysites}a. The twofold rotational symmetry around the $c$-axis forbids the out-of-plane charge polarization and allows only the in-plane polarization along $c$. Using density functional theory (DFT) calculations, we searched for possible symmetry-reduced structures such as an $a$-axis-polar $Cm$ branch (SG \#8). However, these pristine-lattice distortions either forbid an $a$-axis dipole by symmetry or are not energetically favorable. Such a symmetry constraint motivated us to explore alternative causes of the out-of-plane polarization measured by experiment~\cite{wu2026wafer}.

Oxygen vacancies (V$_\text{O}$) are common active defects in oxides and are known to affect charge trapping, ionic migration, and resistive switching~\cite{lee2023role,nukala2021reversible,rushchanskii2021ordering}. To examine the vacancy-driven ferroelectric switching, we first study favored locations and electronic properties of V$_\text{O}$ in Bi$_2$SeO$_5$. Figure~\ref{fig:vacancysites}a indicates the symmetry-inequivalent oxygen sites (1--6) in pristine Bi$_2$SeO$_5$, together with glide-related counterparts (3$'$--5$'$) and an interlayer site across the van der Waals gap (site 7, symmetry-equivalent to site 4). 
The calculated vacancy formation energies as a function of the oxygen chemical potential (Fig.~\ref{fig:vacancysites}b) show that vacancies at sites 3, 4, and 5, which belong to the SeO$_3$ unit, are thermodynamically favored over vacancies inside the Bi$_2$O$_2$ layer (sites 1 and 2) or at the bridging site 6. Thus, V$_\text{O}$ is not distributed uniformly across the lattice but is instead concentrated in a local SeO$_3$ cluster.

For charge-neutral V$_\text{O}$ configurations, all six vacancy sites retain an insulating electronic structure. The vacancy-induced in-gap states are occupied and spatially localized (Fig.~\ref{fig:vacancysites}c--f). The large binding energy of the defect state suppresses electron ionization, and otherwise mobile carriers would screen the ferroelectric polarization. 
Notably, a vacancy at the central site 1 preserves the twofold rotational symmetry, which forbids the out-of-plane dipole. By contrast, the most favorable vacancies at sites 3--5 break this rotational symmetry, allowing the presence of out-of-plane dipoles while preserving the insulating nature of the host compound. 

\begin{figure}
    \centering
    \includegraphics[width=0.95\columnwidth]{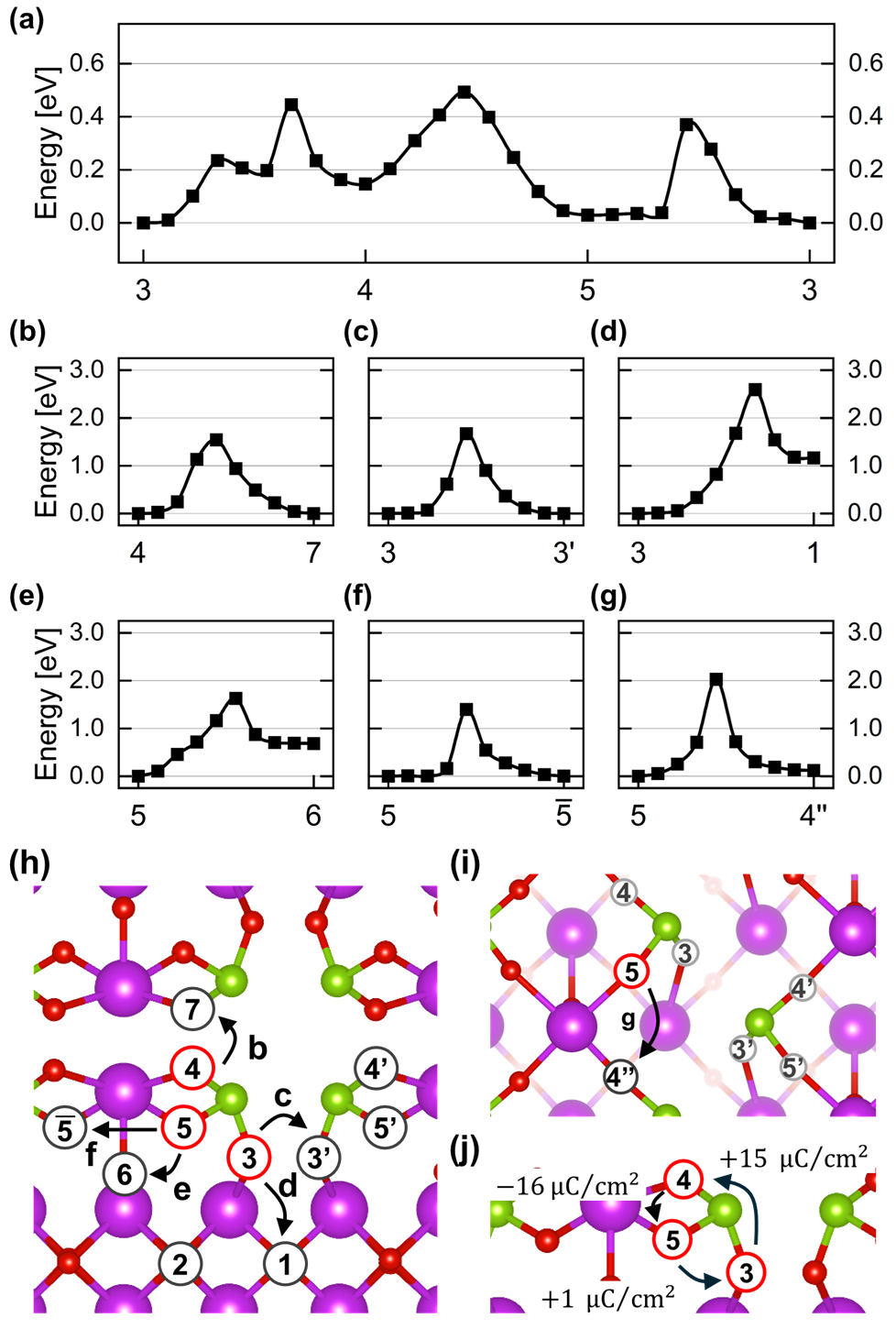}
    \caption{\textbf{Confined vacancy migration barriers and polarization changes.}
    \textbf{a} Low-barrier migration pathway connecting the favored V$_\text{O}$ sites within a SeO$_3$ unit, 3$\rightarrow$4$\rightarrow$5$\rightarrow$3.
    \textbf{b-g} Representative escape pathways from the favored sites illustrated in \textbf{h} and \textbf{i}, showing barriers above 1.5~eV.
    \textbf{h,i} Side and top views of representative escape paths from the SeO$_3$ unit.
    \textbf{j} Confined 3-4-5 rearrangement within the SeO$_3$ unit; moving the vacancy from site 3 or 5 to site 4 changes $P_a$ by about 15--16~$\mu\text{C}\,\text{cm}^{-2}$.
    }
    \label{fig:migration}
\end{figure}

\paragraph*{Energy landscape and confined migration.}

Because these charge-neutral vacancies preferentially occupy sites 3--5, we expect dipole changes induced by vacancy rearrangement to produce switchable ferroelectric polarization. We therefore calculated the out-of-plane polarization change, $\Delta P_{a}^{i\rightarrow j}$, between sites $i$ and $j$, together with the transition barriers along the 3$\rightarrow$4$\rightarrow$5$\rightarrow$3 pathway (see Fig.~\ref{fig:migration}a).

The vacancy migration inside the SeO$_3$ unit shows small barriers. Along the 3$\rightarrow$4$\rightarrow$5$\rightarrow$3 pathway, transition barriers are only $\sim$0.2--0.5~eV (Fig.~\ref{fig:migration}a). 
Here, sites 3 and 5 have similar polarizations with $\Delta P_{a}^{5 \rightarrow 3} = 1~\mu\text{C}\,\text{cm}^{-2}$, while site 4 shows quite distinct polarization from sites 3 and 5, with $\Delta P_{a}^{3 \rightarrow 4} = 15~\mu\text{C}\,\text{cm}^{-2}$ and $\Delta P_{a}^{4 \rightarrow 5} = -16~\mu\text{C}\,\text{cm}^{-2}$. Thus, the relocation from sites 3 or 5 to site 4 is strongly polar-active.

This polar-active motion is confined within the SeO$_3$ unit rather than leading to long-range diffusion. All migration pathways that move the oxygen vacancy out of a local SeO$_3$ unit have large barriers of about 1.5--3.0~eV. As shown in Figs.~\ref{fig:migration}h,i, these escape pathways include relocation from the SeO$_3$ region to the Bi$_2$O$_2$ layer (e.g., 3$\rightarrow$1), across the vdW gap (e.g., 4$\rightarrow$7), in-plane shift along the $b$-axis (e.g., 5$\rightarrow$$\bar{5}$) or other in-plane directions (e.g., 5$\rightarrow $6 or $4''$). The large separation between the local migration barriers and the escape barriers implies kinetic confinement: 
assuming a typical phonon attempt frequency of $\sim$10$^{13}$~Hz, the 0.2--0.5~eV local barriers permit rapid switching (nanoseconds), while escape barriers above 1.5~eV effectively suppress long-range diffusion at room temperature. Thus, the oxygen vacancy is expected to be mobile for local dipole switching, while long-range diffusion is strongly suppressed. This separation between low-barrier local rearrangement and high-barrier long-range migration provides the microscopic basis for switchable vacancy-driven polarization in Bi$_2$SeO$_5$.

\begin{figure}
    \centering
    \includegraphics[width=0.95\columnwidth]{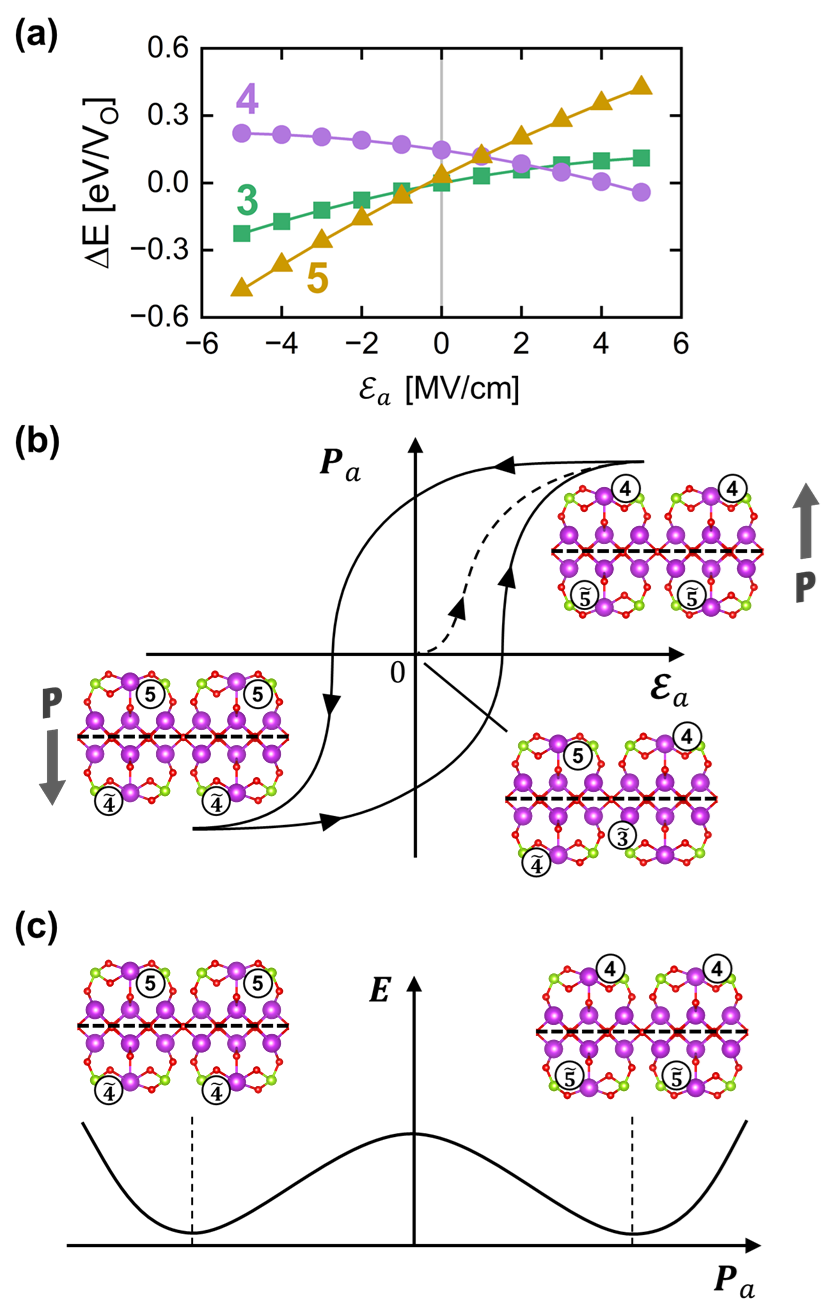}
    \caption{\textbf{Vacancy rearrangement and ferroelectric switching by the electric field.}
    \textbf{a} Relative energies of V$_\text{O}$ at sites 3--5 in the upper half of a Bi$_2$SeO$_5$ layer under an out-of-plane electric field $\mathcal{E}_a$, referenced to site 3. Symbols labeled 3--5 denote the vacancy sites defined in Fig.~\ref{fig:vacancysites}a. Large positive and negative fields favor sites 4 and 5, respectively.
    \textbf{b} Schematic of field-driven V$_\text{O}$ relocation from an initially random distribution over symmetry-related sites, producing the ferroelectric hysteresis.
    \textbf{c} Schematic energy landscape for vacancy-driven polarization switching. Low-barrier V$_\text{O}$ migration within SeO$_3$ units connects the 4/$\tilde{5}$ and 5/$\tilde{4}$ configurations with opposite polarization.
    }
    \label{fig:efield}
\end{figure}

\paragraph*{Polarization switching.}

Next, we demonstrate the controlled migration inside the SeO$_3$ unit by an external electric field. At zero field, sites 3--5 are nearly degenerate in energy, with sites 3 and 5 slightly favored over site 4 (Figs.~\ref{fig:vacancysites}b and \ref{fig:efield}a). An out-of-plane electric field $\mathcal{E}_a$ can reverse the relative order between them. For a vacancy in the upper half of the layer, $+\mathcal{E}_a$ stabilizes site 4, whereas $-\mathcal{E}_a$ stabilizes site 5 (Fig.~\ref{fig:efield}a). This field-induced site selection is the key requirement for a switchable polarization.

This local vacancy site preference induces the macroscopic out-of-plane polarization. To begin with a microscopic perspective, let us first consider a monolayer of Bi$_2$SeO$_5$. Before applying an electric field, oxygen vacancies are expected to populate glide- or mirror-related equivalent sites without a preferred $a$-axis orientation (3 vs $\tilde{3}$, 4 vs $\tilde{4}$, or 5 vs $\tilde{5}$ in Fig.~\ref{fig:efield}b), showing zero net polarization. Because the upper and lower SeO$_3$ units are glide-related, the same out-of-plane field selects different local vacancy sites in the two halves. A strong field $+\mathcal{E}_a$ favors site 4 in the upper SeO$_3$ unit or site $\tilde{5}$ in the lower SeO$_3$ unit, whereas $-\mathcal{E}_a$ favors site $\tilde{4}$ or site 5, leading to opposite net polarizations under $\pm \mathcal{E}_a$. The field required for this site selection is comparable to the experimentally accessible switching-field window of ultrathin Bi$_2$SeO$_5$ ($<$\,8~MV\,cm$^{-1}$)~\cite{wu2026wafer}. For a vacancy concentration of 2.5\%, which corresponds to one vacancy per conventional unit cell, providing a physically meaningful baseline, 
the switchable polarization is $\vert\Delta P_\text{a}^{4\leftrightarrow 5}\vert=16~\mu\text{C}\,\text{cm}^{-2}$, equivalent to the polarization change between sites 4 and 5. This value is comparable to the experimental report of $22~\mu\text{C}\,\text{cm}^{-2}$~\cite{wu2026wafer}. Thus, this field-driven switching defines two oppositely polarized minima connected by low-barrier local migration within each SeO$_3$ cluster, providing a microscopic energy landscape for vacancy-driven switching (Fig.~\ref{fig:efield}c).

The vacancy migration is accompanied by local structural distortion. For the vacancy relocation from site 4 to site 5, our calculated 
cation displacements (0.17~\AA{} for Se and 0.02~\AA{} for Bi along the $a$-axis) agree reasonably with previous experimental measurements (0.29~\AA{} and 0.13~\AA{}, respectively)~\cite{wu2026wafer}.

\paragraph*{Discussion.}

Beyond the dominant switching between the two strongly polar vacancy configurations (sites 4 and 5), the site 3 configuration may also play a functional role as an intermediate or metastable state in the field-driven migration pathway. Although its out-of-plane polarization is close to that of site 5 and therefore may not constitute an independent ferroelectric memory state by itself, its finite stability could still influence switching kinetics and enable multi-step or analog responses under appropriate voltage pulses. This suggests that the vacancy manifold in Bi$_2$SeO$_5$ may be relevant not only to binary ferroelectric switching but also to defect-mediated memristive behavior~\cite{guo2025oxygen,chen2026epitaxial}, which would merit further study.

We should point out that the vacancy-induced ferroelectricity in Bi$_2$SeO$_5$ is different from the case of HfO$_2$. V$_\text{O}$ leads to ferroelectric properties in HfO$_2$ by influencing its polymorphism~\cite{lee2023role,he2021ferroelectric,zhou2019effects,rushchanskii2021ordering}. Because V$_\text{O}$ induces charge-trap states inside the gap, it generates local conduction, which is usually considered harmful for the insulating layer~\cite{gritsenko2016electronic}. It migrates and accumulates to form filaments, leading to the resistive switching mechanisms~\cite{lee2023role,dirkmann2018filament,nukala2021reversible,barneo2026coupling,guo2025oxygen,chen2026epitaxial}. In contrast, V$_\text{O}$ is charge neutral and does not conduct charge in Bi$_2$SeO$_5$. Unlike HfO$_2$, where vacancies form conductive filaments that degrade the dielectric, the deep localized in-gap states and high escape barriers in Bi$_2$SeO$_5$ kinetically and thermodynamically prevent filamentation. This preserves the insulating nature while still enabling ferroelectricity. Here, V$_\text{O}$ directly generates ferroelectric polarization by rearranging within the confined SeO$_3$ unit, without inducing additional bulk polymorphism or charging.

In summary, we propose a vacancy-mediated ferroelectric switching mechanism in Bi$_2$SeO$_5$. Oxygen vacancies preferentially occupy SeO$_3$ units and generate local out-of-plane dipoles. Under an applied electric field, the vacancies switch through low-barrier rearrangements confined within individual SeO$_3$ units. In contrast, long-range vacancy migration has much higher barriers. More broadly, our results suggest a design principle for defect-enabled ferroelectricity beyond the conventional ferroelectric paradigm for a broad family of van der Waals materials. 

\begin{acknowledgments}
We thank Hailin Peng, Huixia Fu, and Hengxin Tan for inspiring discussions. B.Y. acknowledges the financial support by the Israel Science Foundation (ISF: 2974/23), National Science Foundation (NSF DMREF: 2522898), and the Penn State Materials Research Science and Engineering Center for Nanoscale Science (MRSEC) under National Science Foundation award DMR-2011839.
\end{acknowledgments}

\paragraph*{Computational details.}

Spin-polarized density functional theory (DFT) calculations were performed using the Vienna \textit{ab initio} Simulation Package (VASP) as implemented with the projector augmented-wave method~\cite{kresse1999ultrasoft}. The exchange-correlation energy was described by the Perdew-Burke-Ernzerhof (PBE) functional~\cite{perdew1996generalized}, and van der Waals interactions were included using the DFT-D3 method with the Becke-Johnson damping~\cite{grimme2011effect}. For the conventional Bi$_2$SeO$_5$ cell (11.34~\AA{}\,$\times$\,16.47~\AA{}\,$\times$\,5.52~\AA{}), the Brillouin zone was sampled by a $3\times2\times6$ Monkhorst-Pack mesh~\cite{monkhorst1976special}. The electronic energy convergence criterion was $10^{-7}$~eV, and ionic relaxation was performed until the Hellmann-Feynman forces were below 1~meV\,\AA{}$^{-1}$.

Oxygen-vacancy migration pathways were calculated using the climbing-image nudged elastic band (CI-NEB) method~\cite{sheppard2012generalized}. A single oxygen vacancy was introduced into a $1\times1\times2$ supercell to reduce spurious interactions between periodic images. For these defect calculations, the force convergence criterion was relaxed to 10~meV\,\AA{}$^{-1}$.

\paragraph*{Data availability.}
The data that support the findings of this article are available from the corresponding author upon reasonable request.

\bibliography{reference}

@article{kresse1999ultrasoft,
  title={From ultrasoft pseudopotentials to the projector augmented-wave method},
  author={Kresse, Georg and Joubert, Daniel},
  journal={Physical Review B},
  volume={59},
  number={3},
  pages={1758},
  year={1999},
  publisher={American Physical Society}
}

@article{perdew1996generalized,
  title={Generalized gradient approximation made simple},
  author={Perdew, John P and Burke, Kieron and Ernzerhof, Matthias},
  journal={Physical Review Letters},
  volume={77},
  number={18},
  pages={3865},
  year={1996},
  publisher={American Physical Society}
}

@article{grimme2011effect,
  title={Effect of the damping function in dispersion corrected density functional theory},
  author={Grimme, Stefan and Ehrlich, Stephan and Goerigk, Lars},
  journal={Journal of Computational Chemistry},
  volume={32},
  number={7},
  pages={1456--1465},
  year={2011},
  publisher={Wiley Online Library}
}

@article{monkhorst1976special,
  title={Special points for Brillouin-zone integrations},
  author={Monkhorst, Hendrik J and Pack, James D},
  journal={Physical review B},
  volume={13},
  number={12},
  pages={5188},
  year={1976},
  publisher={APS}
}

@article{sheppard2012generalized,
  title={A generalized solid-state nudged elastic band method},
  author={Sheppard, Daniel and Xiao, Penghao and Chemelewski, William and Johnson, Duane D and Henkelman, Graeme},
  journal={The Journal of chemical physics},
  volume={136},
  number={7},
  year={2012},
  publisher={AIP Publishing}
}

@article{cheema2020enhanced,
  title={Enhanced ferroelectricity in ultrathin films grown directly on silicon},
  author={Cheema, Suraj S and Kwon, Daewoong and Shanker, Nirmaan and Dos Reis, Roberto and Hsu, Shang-Lin and Xiao, Jun and Zhang, Haigang and Wagner, Ryan and Datar, Adhiraj and McCarter, Margaret R and others},
  journal={Nature},
  volume={580},
  number={7804},
  pages={478--482},
  year={2020},
  publisher={Nature Publishing Group UK London}
}

@article{cheema2022emergent,
  title={Emergent ferroelectricity in subnanometer binary oxide films on silicon},
  author={Cheema, Suraj S and Shanker, Nirmaan and Hsu, Shang-Lin and Rho, Yoonsoo and Hsu, Cheng-Hsiang and Stoica, Vladimir A and Zhang, Zhan and Freeland, John W and Shafer, Padraic and Grigoropoulos, Costas P and others},
  journal={Science},
  volume={376},
  number={6593},
  pages={648--652},
  year={2022},
  publisher={American Association for the Advancement of Science}
}

@article{khan2020future,
  title={The future of ferroelectric field-effect transistor technology},
  author={Khan, Asif Islam and Keshavarzi, Ali and Datta, Suman},
  journal={Nature Electronics},
  volume={3},
  number={10},
  pages={588--597},
  year={2020},
  publisher={Nature Publishing Group UK London}
}

@article{nukala2021reversible,
  title={Reversible oxygen migration and phase transitions in hafnia-based ferroelectric devices},
  author={Nukala, Pavan and Ahmadi, Majid and Wei, Yingfen and De Graaf, Sytze and Stylianidis, Evgenios and Chakrabortty, Tuhin and Matzen, Sylvia and Zandbergen, Henny W and Bj{\"o}rling, Alexander and Mannix, Dan and others},
  journal={Science},
  volume={372},
  number={6542},
  pages={630--635},
  year={2021},
  publisher={American Association for the Advancement of Science}
}

@article{taniguchi2013ferroelectricity,
  title={Ferroelectricity driven by twisting of silicate tetrahedral chains},
  author={Taniguchi, Hiroki and Kuwabara, Akihide and Kim, Jungeun and Kim, Younghun and Moriwake, Hiroki and Kim, Sungwng and Hoshiyama, Takuya and Koyama, Tsukasa and Mori, Shigeo and Takata, Masaki and Hosono, Hideo and Inaguma, Yoshiyuki and Itoh, Mitsuru},
  journal={Angewandte Chemie International Edition},
  volume={52},
  number={31},
  pages={8088--8092},
  year={2013},
  publisher={Wiley Online Library}
}

@article{wang2023towards,
  title={Towards two-dimensional van der Waals ferroelectrics},
  author={Wang, Chuanshou and You, Lu and Cobden, David and Wang, Junling},
  journal={Nature Materials},
  volume={22},
  number={5},
  pages={542--552},
  year={2023},
  publisher={Nature Publishing Group UK London}
}

@article{xue2021emerging,
  title={Emerging van der Waals ferroelectrics: Unique properties and novel devices},
  author={Xue, Fei and He, Jr-Hau and Zhang, Xixiang},
  journal={Applied Physics Reviews},
  volume={8},
  number={2},
  year={2021},
  publisher={AIP Publishing}
}

@article{zhou2019effects,
  title={The effects of oxygen vacancies on ferroelectric phase transition of HfO2-based thin film from first-principle},
  author={Zhou, Ying and Zhang, YK and Yang, Qiong and Jiang, Jie and Fan, Pan and Liao, Min and Zhou, YC},
  journal={Computational Materials Science},
  volume={167},
  pages={143--150},
  year={2019},
  publisher={Elsevier}
}

@article{he2021ferroelectric,
  title={Ferroelectric structural transition in hafnium oxide induced by charged oxygen vacancies},
  author={He, Ri and Wu, Hongyu and Liu, Shi and Liu, Houfang and Zhong, Zhicheng},
  journal={Physical Review B},
  volume={104},
  number={18},
  pages={L180102},
  year={2021},
  publisher={APS}
}

@article{lee2023role,
  title={Role of oxygen vacancies in ferroelectric or resistive switching hafnium oxide},
  author={Lee, Jaewook and Yang, Kun and Kwon, Ju Young and Kim, Ji Eun and Han, Dong In and Lee, Dong Hyun and Yoon, Jung Ho and Park, Min Hyuk},
  journal={Nano Convergence},
  volume={10},
  number={1},
  pages={55},
  year={2023},
  publisher={Springer}
}

@article{rushchanskii2021ordering,
  title={Ordering of oxygen vacancies and related ferroelectric properties in HfO 2-$\delta$},
  author={Rushchanskii, Konstantin Z and Bl{\"u}gel, Stefan and Le{\v{z}}ai{\'c}, Marjana},
  journal={Physical review letters},
  volume={127},
  number={8},
  pages={087602},
  year={2021},
  publisher={APS}
}

@article{li2020native,
  title={A native oxide high-$\kappa$ gate dielectric for two-dimensional electronics},
  author={Li, Tianran and Tu, Teng and Sun, Yuanwei and Fu, Huixia and Yu, Jia and Xing, Lei and Wang, Ziang and Wang, Huimin and Jia, Rundong and Wu, Jinxiong and others},
  journal={Nature Electronics},
  volume={3},
  number={8},
  pages={473--478},
  year={2020},
  publisher={Nature Publishing Group UK London}
}

@article{dong2024exploring,
  title={Exploring the high dielectric performance of Bi2SeO5: from bulk to bilayer and monolayer},
  author={Dong, Xinyue and He, Yuyu and Guan, Yue and Zhu, Yuanhao and Wu, Jinxiong and Fu, Huixia and Yan, Binghai},
  journal={Science China Materials},
  volume={67},
  number={3},
  pages={906--913},
  year={2024},
  publisher={Springer}
}

@article{zhang2023single,
  title={Single-crystalline van der Waals layered dielectric with high dielectric constant},
  author={Zhang, Congcong and Tu, Teng and Wang, Jingyue and Zhu, Yongchao and Tan, Congwei and Chen, Liang and Wu, Mei and Zhu, Ruixue and Liu, Yizhou and Fu, Huixia and others},
  journal={Nature materials},
  volume={22},
  number={7},
  pages={832--837},
  year={2023},
  publisher={Nature Publishing Group UK London}
}

@article{wu2026wafer,
  title={Wafer-scale ultrathin and uniform van der Waals ferroelectric oxide},
  author={Wu, Qinci and Li, Zhongrui and Han, Bingchen and Sun, Weiyu and Liu, Qinyun and Xue, Chengyuan and Bae, Hyeonhu and Wang, Mengdi and Fu, Boyang and Qian, Jun and others},
  journal={Science},
  volume={391},
  number={6784},
  pages={eadz1655},
  year={2026},
  publisher={American Association for the Advancement of Science}
}

@article{guo2025oxygen,
  title={Oxygen Vacancy Induced 2D Bi2SeO5 Non-Volatile Memristor for 1T1R Integration},
  author={Guo, Tingting and Pan, Zhidong and Shen, Yehui and Yang, Jialin and Chen, Chuyao and Xiong, Yunhai and Chen, Xuan and Song, Yang and Huo, Nengjie and Xu, Rongqing and others},
  journal={Nano Letters},
  volume={25},
  number={20},
  pages={8258--8266},
  year={2025},
  publisher={ACS Publications}
}

@article{gritsenko2016electronic,
  title={Electronic properties of hafnium oxide: A contribution from defects and traps},
  author={Gritsenko, Vladimir A and Perevalov, Timofey V and Islamov, Damir R},
  journal={Physics Reports},
  volume={613},
  pages={1--20},
  year={2016},
  publisher={Elsevier}
}

@article{dirkmann2018filament,
  title={Filament growth and resistive switching in hafnium oxide memristive devices},
  author={Dirkmann, Sven and Kaiser, Jan and Wenger, Christian and Mussenbrock, Thomas},
  journal={ACS applied materials \& interfaces},
  volume={10},
  number={17},
  pages={14857--14868},
  year={2018},
  publisher={ACS Publications}
}

@article{barneo2026coupling,
  title={The Coupling of Ferroelectric Polarization and Oxygen Vacancy Migration Enables Electrically Controlled Thermal Memories},
  author={Barneo, D{\'\i}dac and Royo, Miquel and Ramos, Rafael and Carrete, Jes{\'u}s and Romero-Bernad, Hugo and Jim{\'e}nez, Ricardo and Lebor{\'a}n, V{\'\i}ctor and Mag{\'e}n, C{\'e}sar and Varela-Dom{\'\i}nguez, Noa and Alguer{\'o}, Miguel and others},
  journal={Advanced Materials},
  volume={38},
  number={24},
  pages={e19670},
  year={2026},
  publisher={Wiley Online Library}
}

@article{dawber2005physics,
  title={Physics of thin-film ferroelectric oxides},
  author={Dawber, Matthew and Rabe, KM and Scott, JF},
  journal={Reviews of modern physics},
  volume={77},
  number={4},
  pages={1083--1130},
  year={2005},
  publisher={APS}
}

@article{rademacher2001crystal,
  title={Crystal structure of dibismuth selenium pentoxide, Bi2SeO5},
  author={Rademacher, O and G{\"o}bel, H and Ruck, M and Oppermann, H},
  journal={Zeitschrift f{\"u}r Kristallographie - New Crystal Structures},
  volume={216},
  number={1-4},
  pages={29--30},
  year={2001},
  publisher={OLDENBOURG WISSENSCHAFTSVERLAG}
}

@article{dityatyev2004phase,
  title={Phase equilibria in the Bi2TeO5-Bi2SeO5 system and a high temperature neutron powder diffraction study of Bi2SeO5},
  author={Dityatyev, Oleg A and Smidt, Peer and Stefanovich, Sergey Yu and Lightfoot, Philip and Dolgikh, Valery A and Opperman, Heinrich},
  journal={Solid state sciences},
  volume={6},
  number={9},
  pages={915--922},
  year={2004},
  publisher={Elsevier}
}

@article{chen2026epitaxial,
  title={Epitaxial Bi2O2Se/Bi2O5Se Thin Films: Revealing Electric-Field-Driven Oxidation and Resistive Switching Dynamics for Advanced Memory Devices},
  author={Chen, Yen-Jung and Wang, Yong-Jyun and Hong, Zi-Qin and Wang, Chien-Hua and Wang, Yi-Ning and De Cheng, Jia and Huang, Chun-Wei and Chu, Ying-Hao and Wu, Wen-Wei},
  journal={Advanced Science},
  pages={e75508},
  year={2026},
  publisher={Wiley Online Library}
}

\end{document}